\documentclass{article}

\usepackage{arxiv}

\usepackage[utf8]{inputenc} 
\usepackage[T1]{fontenc}    
\usepackage{hyperref}       
\usepackage{url}            
\usepackage{booktabs}       
\usepackage{amsfonts}       
\usepackage{nicefrac}       
\usepackage{microtype}      
\usepackage{lipsum}		
\usepackage{doi}

\usepackage{graphicx}

\usepackage{floatrow}
\newfloatcommand{capbtabbox}{table}[][\FBwidth]
\usepackage{blindtext}

\title{AUTOMATIC DETECTION OF NOISE EVENTS\\ AT SHOOTING RANGE USING MACHINE LEARNING}

\author{
    Jon Nordby \\
	Soundsensing AS \\
	\And
	Fabian Nemazi \\
	Norwegian University of Life Sciences \\
	\And
    Dag Rieber \\
	Rieber Prosjekt AS \\
}

\date{}

\begin{document}

\maketitle
\renewcommand{\abstractname}{\vspace{-\baselineskip}} 

\begin{abstract}	\noindent
Outdoor shooting ranges are subject to noise regulations from local and national authorities. Restrictions found in these regulations may include limits on times of activities,
the overall number of noise events, as well as limits on number of events depending on the class of noise or activity.
A noise monitoring system may be used to track overall sound levels,
but rarely provide the ability to detect activity or count the number of events,
required to compare directly with such regulations.
This work investigates the feasibility and performance of an automatic detection system to count noise events.
An empirical evaluation was done by collecting data at a newly constructed shooting range and training facility.
The data includes tests of multiple weapon configurations from small firearms to high caliber rifles and explosives, at multiple source positions, and collected on multiple different days.
Several alternative machine learning models are tested, using as inputs time-series of standard acoustic indicators such as A-weighted sound levels and 1/3 octave spectrogram, and classifiers such as Logistic Regression and Convolutional Neural Networks.
Performance for the various alternatives are reported in terms of the False Positive Rate and False Negative Rate.
The detection performance was found to be satisfactory for use in automatic logging of time-periods with training activity. \\

\end{abstract}

\quad\rule{425pt}{0.4pt}

\section{Introduction}


Outdoor shooting ranges and police training facilities are subject to noise regulations from local and national authorities,
and site operators must document adherence to these regulations.
Some aspects, such as how much noise is propagated outwards from each noise event,
may be adequately tested upon completion of construction.
Other regulations, such as what time of day noise is emitted or limits on the number of noise-emitting activities, depend more on the ongoing operations.
By using continuous noise monitoring with a sensor network, one can also document these.
A standard noise monitoring sensor provides sound level logging over time, but no direct indicators for the activities.
In this work, a noise monitoring sensor system is described that has the ability to
automatically detect and count noise events from gunshots and explosions.
The system has been developed and tested at the newly constructed Police National Emergency Response Center (Norwegian: \textit{Politiets Nasjonale Beredskapssenter}, abbreviated: PNB) near Oslo, Norway.
The solution has been developed by Soundsensing in collaboration with PNB.

\clearpage
\section{Background}

\subsection{Politiets Nasjonale Beredskapssenter}

Politiets Nasjonale Beredskapssenter (PNB) is a combined training facility and operative base for the police special forces in Norway.
It consists of an office building, an airport for police helicopters
and training facilities with indoor and outdoor shooting ranges for rifles and handguns,
as well as buildings for practicing breach and entry with explosives.

A key criteria for the location of the center was proximity to Oslo city and urban area,
in order to ensure fast response times.
Another criteria was a location far from residential areas, to prevent disturbance caused by noise from shooting and explosives used during training.
The chosen location combines these two requirements:
It is within 20 minutes drive to downtown Oslo, and noise is managed by extensive noise abatement measures\cite{RieberPNBForbedredeTiltak}. Some examples of the constructed measures:
Training areas are surrounded by 10 meter tall berms, with additional 2 meter tall
noise dampening fences.
For the outdoor shooting ranges the shooting positions are inside a building that is closed on all sides apart from the shooting direction.
Buildings used for explosives training are fitted with noise absorbing walls and interiors.


\subsection{Regulations}

In Norway the recommended noise limits for shooting-ranges and other kinds of environmental noise is defined in T-1442/2016\cite{T-1442/2016}.
It defines two zones, a red zone which is not suitable for noise sensitive buildings,
and a yellow zone where noise sensitive buildings may be placed only if measures are made to provide 
acceptable noise conditions.  
For shooting-ranges the red zone is defined by $L_{den}$\cite{EuNoiseDirective} of 35 dB and $L_{AFmax}$\cite{IECSoundLevelMeters} of 65 dB,
and the yellow zone is defined by $L_{den}$ 45 dB and $L_{AFmax}$ 75 dB.
At night (between 23.00 and 07.00) no activity should be allowed.

The local municipality use these recommendations as a starting point,
and define the local regulations though a zoning plan\cite{ZoningPlanNorway}.
In the case of PNB, concern for noise from the local residents was very high.
For this reason the authorities set much stricter regulations\cite{pnb-reguleringsplan} regarding noise from training activities than the recommended noise limits found in T-1442/2016.
The additional restrictions include:

\begin{enumerate}
  \item Training activity is only permitted on Monday - Friday, between 07:00 and 19:00.
  \item The number of explosive charges outdoors may not exceed 1250 per year.
  \item Maximum number of gunshots per year of 2000 0000.
  \item The yellow and red zone may not exceed the original approved plans.
\end{enumerate}

\subsection{Noise characteristics}

Shooting range noise usually has three different sound generation mechanisms: The muzzle blast, the bullet's supersonic boom and impact noise from targets and bullet traps.
The muzzle blast comes from combustion gas expelled from the barrel of the firearm, caused by the combustion of gunpowder.
It expands rapidly and causes a sound wave. With a shorter barrel normally more of the gunpowder is combusted outside the barrel, causing louder noise.
When the speed of the bullet exceeds the speed of sound, a supersonic boom is created.
The supersonic boom travels in an angle from the bullet’s trajectory. The boom varies by the bullet speed, the bullet length, diameter and shape.

Shooting at steel targets and steel bullet-traps can cause severe impact noise.
To limit pollution from the shooting ranges to a minimum, targets and bullet traps that fragment the bullets will not be used at PNB.
Therefore, this kind of impact noise will not occur.

Penthrite (Pentaerythritol tetranitrate) is an explosive used for door and window breaching training at the center. A detonation of Penthrite creates a supersonic shockwave that results in a soundwave some milliseconds after the detonation. The sound created by an explosion is heard as one single boom.

The on-site helicopter port will also generate occasional noise.
Activity from helicopter traffic was considered well monitored though
the aviation logbooks, and is thus not important to track using the noise detection system.
The site is located close to a highway with considerable amount of traffic. This is the main source of environmental noise from other sources than PNB. The passing of cars and helicopters has a quite different characteristic than gunshots and explosives, but constitutes a time-varying background noise. 
Construction activities at the center may be a source of other impulse sounds that may be confused with training activity.

\subsection{Related work}

The general task of detecting potentially noisy sounds from the environment (\emph{Environmental Sound Classification}) is well established in the machine learning community, with several datasets \cite{piczak2015dataset:esc50} \cite{salamon_dataset_2014:urbansound8k} \cite{cartwright_sonyc-ust-v2_2020} and proposed methods based on many variations of machine learning \cite{salamon_deep_2017:SBCNN} \cite{cramer_look_2019:OpenL3} \cite{zhang_deep_2018} \cite{abdoli_end2end_2019}.


Some works have explored privacy-aware noise classification with wireless sensor network, either by transmitting specially-designed sound representations  \cite{gontier_efficient_2017:censecoder},
or by doing classification on-sensor  \cite{maijala_environmental_2018}\cite{nordby_environmental_2019}.

The task of detecting short sound events (\emph{Sound Event Detection}) has also received considerable attention. It has features in several challenges as part of the Detection and Classification of Audio Scenes and Events (DCASE) workshop series \cite{stowell_detection_2015}\cite{lafay_sound_2017}\cite{mesaros_sound_2019}.
The sound event detection principle has been used to create better models for road-noise measurements \cite{socoro_anomalous_2017}.

Gunshot detection is also a widely studied problem in the context of security systems \cite{valenzise_scream_2007}.
However, the authors have not been able to find automatic detection of gunshots and explosions used for the purpose of noise monitoring.


\section{Methods}

\subsection{Data collection}

Data collection for the noise detection system was done during tests conducted to verify the effectiveness of the noise reduction measures, and that noise propagation did not exceed limits set in the zoning plan.
The tests followed the guidelines in M128/2014, chapter 9.5.2 \cite{Veileder-M128/2014}.
Each test consisted of either 5 or 10 repetitions with the same weapon configuration and location, performed by a single user at a fixed 30-second interval.
The tests were performed over 4 separate days in the period May to September.
Two days was for testing noise propagation in northward direction,
and two days for noise propagation westward.

\begin{figure}[!h]
\centering
\includegraphics[width=17.0cm,height=6.0cm]{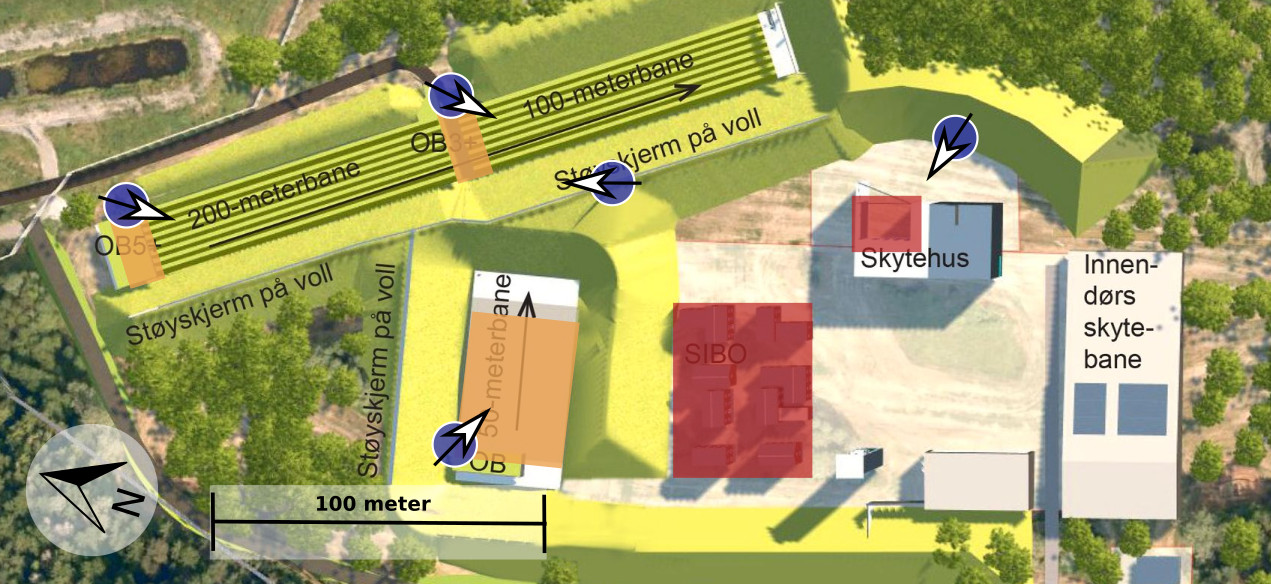}
\caption{ Training facilities, with noise sources and recording locations.  Yellow: Positions of gunshot noise sources. Red: Positions of flashbang and explosives noise sources. Blue: Position and orientation of recording equipment. }
\label{fig:data-collect-center-locations}
\end{figure}

Sound was collected from 64 tests, for a total of 510 noise events.
130 events were explosives (95 flashbang and 35 explosive charges) at the two areas with buildings for explosives training.
The remaining 380 events were single shots from 6 different weapon configurations at the 3 outdoor shooting ranges.
5 audio recorders (with 2 channels each) were used on-site to capture the sound.
The placements can be seen in Figure \ref{fig:data-collect-center-locations}.

Sound was also simultaneously recorded with 6 audio recorders in the surrounding area,
however this data was not used in this work.
In total around 300 hours of sound was recorded for analysis.
Under 1\% contains a noise event of interest,
the rest is background noise from the environment (car traffic on highway, construction work on site and outside, footfall etc.).

Time-stamps labels for the noise events were made semi-automatically.
First a Hidden Markov Model \cite{rabiner1986:HMM} with 2 states was used on the A-weighted sound levels \cite{IECSoundLevelMeters} to automatically detect a probable time of noise events.
This was then visually confirmed and cleaned up, and then verified by listening to the audio.
Time-stamps of the events are estimated to have a precision of within 0.1 seconds.


\subsection{Synthesized dataset}

Since the original data was captured during acoustics testing before start of training activities, it does not reflect planned operating conditions in several regards.
To improve this a new dataset was made by combining recorded noise events and backgrounds to form new synthetic mixtures that are closer to operating conditions,
using the Scaper \cite{salamon_scaper_2017} tool.

The occurrence of events were modelled using a conditional Poisson distribution,
with a constant rate inside the active hours, and a 0 rate outside active hours.
The number of active hours in a year was estimated at 1260 hours.
With 2 million gunshots and 1250 explosives per year (the maximum allowed in regulations), this corresponds to event rates of 1587 gunshots/hour and 0.98 explosions/hour.
To get a number of events with explosives sufficient to estimate performance,
the event rate was artificially increased to 100x the normal.
The distribution of signal-to-noise ratio was selected to be slightly lower than found in recordings.
Each generated mixture was 20 seconds long, and 10000 such mixture were generated.
This results in a dataset which reflects the key characteristics of normal operations; with varying background noise, variable time between events, occasional overlapping events, combinations of explosions and gunshots, representative loudness of events and number of events versus no-event.

\subsection{Sound Event Detection using Machine Learning}

Learning a model to detect noise events can be treated as a supervised classification problem, using the annotated time-period with events as the targets.
The task considered is that of Sound Event Detection of noise events: Any noise event from training activity (gunshot or explosion) versus no event (background / other noises).
The audio stream is divided into overlapping windows of fixed duration,
with 1 second window length and 0.125 seconds hop between each window.
Windows that contain \emph{at least one} noise event of interest is given a positive label, and other windows are given a negative label.

One tested model used a set of hand-designed features derived from A-weighted sound levels with "Fast"\cite{IECSoundLevelMeters} integration time, using either Logistic Regression or Random Forest as the classifier.
These features capture coarse aspects of the impulsive character of a sounds,
but does not have the complete temporal envelope or any information about the spectral content. The features included were:

\begin{enumerate}
  \item Max sound level
  \item Max difference in sound level from the median
  \item Max positive change in sound level between consecutive time steps
  \item Max negative change in sound level between consecutive time steps
  \item Time difference between max negative and max positive change 
\end{enumerate}

Another model used 1/3 octave \cite{IECOctaveBands} spectrograms with 125 ms hop length. The low temporal resolution is chosen such that speech cannot be recovered \cite{gontier_efficient_2017:censecoder}.
For classifying the spectrogram a small Convolutional Neural Network was used.
Such a combination has the ability to learn complex time-frequency patterns \cite{salamon_deep_2017:SBCNN} \cite{nordby_environmental_2019}.
The input to the network is spectrogram window of 27 frequency bands, and 8 time-steps.
The network consists of two convolutional layers, with max pooling after each layer, followed by two fully-connected layers with ReLu activation, and a final fully-connected layer with sigmoid activation.
Both the sound level and spectrogram representations are supported by the sensor nodes that will be used (\emph{Soundsensing dB20}).

\section{Results}

Initial data exploration showed that all events from all locations (shooting ranges and explosive training locations) could be clearly heard at the device at the center of the site.
Therefore results are reported using data only from this device, to evaluate if adequate performance could be achieved without needing more sensors.

\begin{figure}[hbt!]
  \includegraphics[width=17.0cm,height=3.5cm]{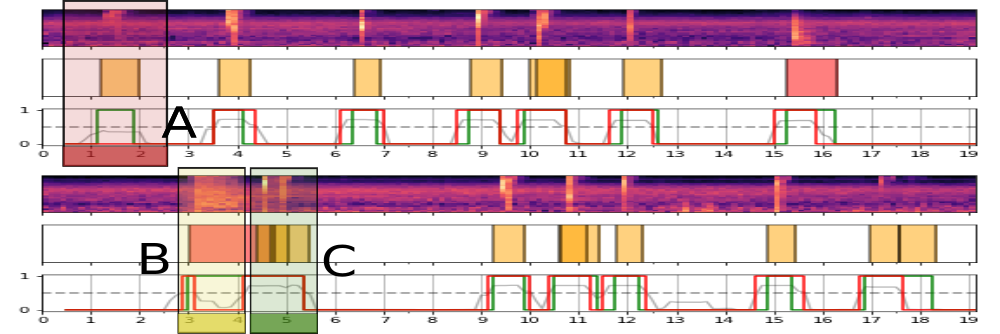}
  \caption{Two timelines with input spectrogram (top), ground truth event labels (middle) and model decisions (bottom, red line). Scenario A: missing detection. Scenario B: Detection of explosion event, decay below threshold. Scenario C: Multiple gunshot events merged into one detected noise event.
  Model used was Random Forest.
  }
  \label{fig:detection-timeline}
\end{figure}

The model can analyze an incoming stream of audio data, and the output is a time-series with the estimated probability of noise events occurring. A threshold can be applied in order to get a binary event vs no-event detection. This usage is shown in \ref{fig:detection-timeline}. In a deployment scenario, the conversion from audio to the sound levels and spectrograms used by models, happens on the sensor.

\begin{figure}[hbt!]
\begin{floatrow}
\ffigbox[.45\textwidth]{%
  \includegraphics[width=7.0cm, height=3.5cm]{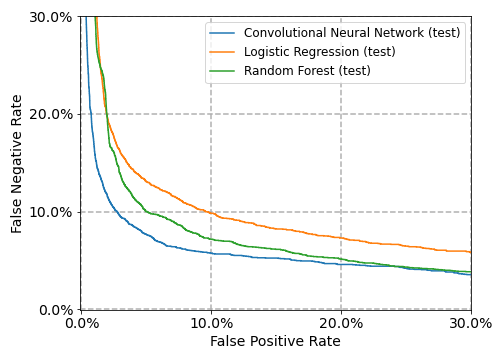}
}{%
  \caption{Detection Error Tradeoff curve for the tested models, on individual windows.}%
  \label{fig:detection-error-tradeoff}
}
\capbtabbox[.35\textwidth]{%
    \begin{tabular}{lrr}
\toprule
Model &   F1 &   AP \\
\midrule
Convolutional Neural Network   & 76.4 & 85.1 \\
Logistic Regression    & 54.3 & 66.8 \\
Random Forest    & 49.0 & 68.2 \\
\bottomrule
\end{tabular}
}{%
  \caption{Evaluation metrics, on individual windows. On the held-out test set }%
  \label{table:performance-metrics}
}
\end{floatrow}
\end{figure}

The model performance was estimated both using per-window evaluation,
with results are reported in \ref{table:performance-metrics}.
The best model was evaluated with per-event metrics (using sed-eval \cite{mesaros_sed_eval}).
It obtained an event-wise
F1 score of 80.5\%, with a precision of 95.5\% and recall of 69.6\% (insertion error rate of 4.5\% and deletion error rate of 30.4\%).


\section{Discussion}

\begin{figure}[hbt!]
  \includegraphics[width=17.0cm, height=2.0cm]{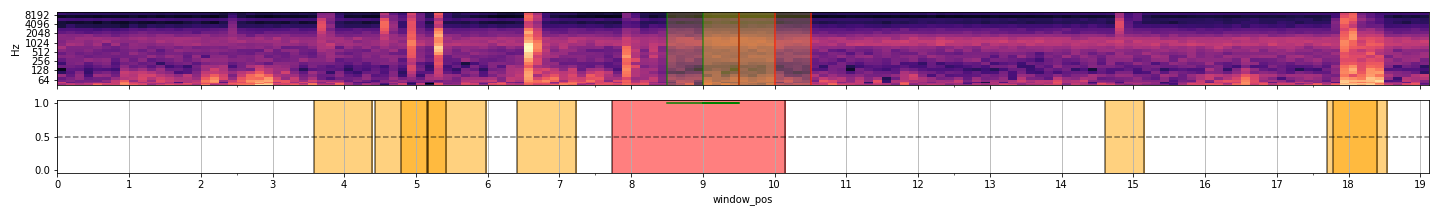}
  \caption{Timeline showing False Negative error with input spectrogram (top), ground truth event labels (bottom). The ground truth suggests that a noise event has occurred, but in fact the impulsive sound of the event onset is not present in this window.
  }
  \label{fig:false-negatives}
\end{figure}


Several of the models are able to classify individual windows with a combined False Positive Rate and False Negative rate of under 10\%.
The Convolutional Neural Network model obtained the best results.
It has a considerably better False Positive Rate over the other two models,
which is important due to the high ratio of time-periods without noise events compared to with noise events.

An analysis of false negatives and false positive showed that a considerable amount of windows were actually classified correctly, but that the labels were incorrect. An example of this is shown in Figure \ref{fig:false-negatives}.
This indicates that it may be necessary to restrict event labels to only cover the start of the event (which contains the impulsive part of the sound) - and that this may improve performance.

One goal in the project is to create a log-book for the training activities that create noise.
Because training activities always consisting of a number of noise events clustered together, this can be achieved even though there are some events missed.

Another goal was to be able to estimate the number of events to within -+10\%.
The insertion error (false events) of 4.5\% is within this requirement, but a deletion rate (missed events) of 30.5\% is not. It may however be that improving the labels (as mentioned in previous section), will allow to reach that performance level.





\section{Further work}

The next step in the project is to install sensors permanently on site,
and validate the estimated performance under regular operating conditions.
A logbook will be generated from the automated event detection,
and will be used to document that activities are in compliance with regulations,
and as data source when handling complaints and requests for information.

It is also planned to survey the noise in the surrounding area during first the year of operation.
The automatic event detection will be used together with noise monitoring sensors near residential areas, to estimate the amount of noise that originates from activities at the training facility versus other sources.

We are also looking to test the developed system at other shooting ranges and training facilities, as well as other use-cases that contain impulsive noise, such as industrial and construction sites.


\section*{ACKNOWLEDGEMENTS}

The work has been funded though a collaboration between Politiets Nasjonale Beredskapsenter construction project and Soundsensing AS.
In the first months of the project, Soundsensing received funding from the Research Council of Norway as part of the program STUD-ENT.

Organization of acoustical tests was lead by Lars R. Nordin at Brekke \& Strand AS.
Data collection for Soundsensing AS was performed by Erik Sjølund and Ole Johan Aspestrand Bjerke.
We thank the entire team for excellent facilitation and execution.


\bibliographystyle{ieeetr}
\bibliography{main}

\end{document}